\documentclass[12pt]{iopart}
\usepackage{iopams}

\expandafter\let\csname equation*\endcsname\relax

\expandafter\let\csname endequation*\endcsname\relax
\usepackage{subfig}
\usepackage{amsmath}
\usepackage{multicol}
\usepackage{subfig}
\usepackage[numbers]{natbib}
\setcitestyle{open={[},close={]}}

\newcommand{\cch}[1]{\left[#1\right]}
\newcommand{\cha}[1]{\left\{#1\right\}}
\newcommand{\prt}[1]{\left(#1\right)}

\usepackage[colorinlistoftodos]{todonotes}
\usepackage[hidelinks,colorlinks,breaklinks=true]{hyperref}
\hypersetup{
  colorlinks=true, linkcolor=blue, 
  citecolor=blue,   
  filecolor=magenta,
  urlcolor=blue,
  bookmarksdepth=4
}

\begin{document}

\title{Coulomb interaction rules timescales in potassium ion channel tunneling}

\author{N. De March, S. D. Prado and L. G. Brunnet} 

\address{Instituto de F\'{\i}sica, Universidade Federal do Rio Grande do Sul (UFRGS) \\
CP: 15051 Porto Alegre, RS, Brazil}
\ead{nicole.march@ufrgs.br}
\vspace{10pt}
\begin{indented}
\item[]\today
\end{indented}

\begin{abstract}
Assuming the selectivity filter of KcsA potassium ion channel may exhibit quantum coherence, we extend a previous  model by Vaziri \& Plenio \cite{vaziri2010quantum} to  take into account Coulomb repulsion between potassium ions. We show that typical ion transit timescales are determined by this interaction, which imposes optimal input/output parameter ranges. Also, as observed in other examples of  quantum tunneling in biological systems, addition of moderate noise helps coherent ion transport. 
\end{abstract}


\vspace{2pc}

\submitto{\JPCM}

\maketitle

%

\section{Introduction}

Ion channels are structures formed by  protein complexes which regulate ion flow through cell membrane.
 They have a filter which selects a specific ion and  transfers it very efficiently. This selectivity filter (SF) is a protein structural motif forming a tunnel inside the ion channel. 
Potassium channels are the most studied types of ion channels since they are responsible for both establishing the resting potential in most cells and the asymptotic shape of the action potential in excitable cells.

\begin{figure}[ht]
\centering
\includegraphics[width=0.5\linewidth,clip=true]{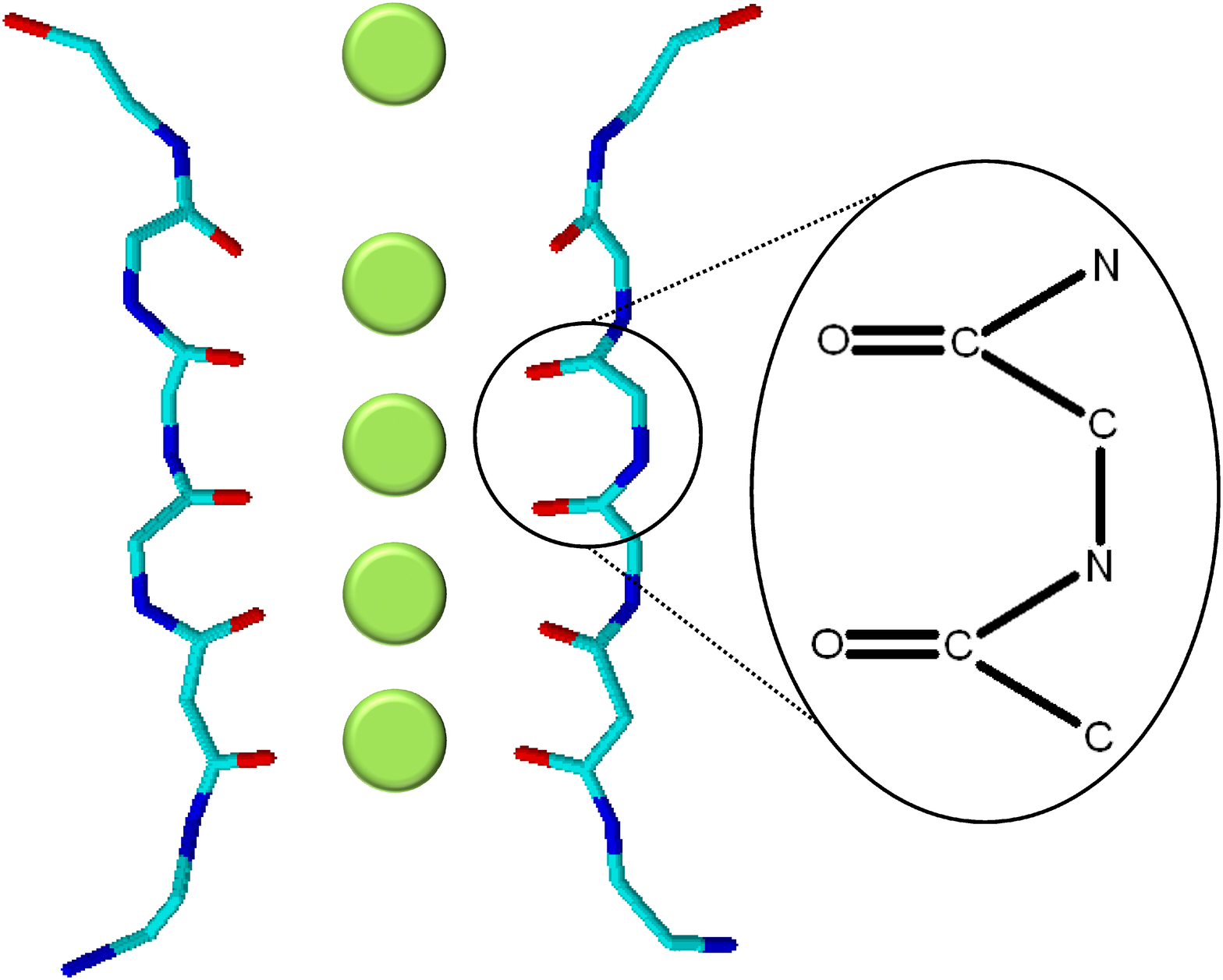}
\caption{KcsA selectivity filter with four axial binding sites formed by peptide units H-N-C=O which carbonyl oxygen atoms C=O point toward the pore trapping a potassium ion or a water molecule.}
\label{Kanal}
\end{figure}

The  first crystallized potassium ion-channel KcsA of the bacteria {\it Streptomyces lividans} \cite{doyle1998structure} has provided a better understanding on the SF  structure. The KcsA filter is about 1.2 nm long with a diameter of 0.3 nm. It is believed that due to this  narrowness and the fact that the incoming ions are without their hydration  shell \cite{roux2004computational},  the oxygen atoms of the  carbonyl groups coordinate the ions motion into a single file fashion (Fig. \ref{Kanal}). This is the fundamental structure under the four existing binding sites forming the ion path. 
While ion-channels structure may differ from bacterial to  eukaryotic cells, the KcsA type  remains relevant given that the SF structure is almost the same across species. One of the most remarkable feature is the high current of $10^{8}$ ions per second \cite{gouaux2005principles} which is comparable to free-ion diffusion in water \cite{hille2001ion}.

 Classical models have been proposed to  explain the KcsA high throughput rate, and the two major ones are cotranslocations of ions with water and direct knock-on. 
Ion translocation in KcsA is constructed considering states that alternate potassium ions e water molecules inside the filter. 
The main argument for cotranslocation relies on the large  electrostatic repulsion between  ions in adjacent sites, making it energetically  unfavorable for two ions to occupy neighboring sites \cite{morais2001energetic,zhou2003occupancy}. So, although  it would be possible to have ions side by side, given their size, it was originally assumed two ions typically separated by a single water molecule, the water being  then responsible for the shielded Coulomb repulsion \cite{morais2001energetic}. Besides that, the knock-on mechanism assumes that a third incoming ion interacts with two ions already inside the filter pushing them towards the flux direction driving the ion flow \cite{hodgkin1955potassium,berneche2001energetics}. 
 Computational simulations \cite{furini2009atypical,fowler2013energetics} compare  configurations with water (KWK) in the channel and with ions in adjacent sites (KK) opening the possibility of side by side configuration for potassium ions.  Some authors  \cite{bordin2012ion,kopfer2014ion} explore this point further focusing on  Coulomb repulsion as the main ingredient driving the process efficiency.

Inspired by the recent achievements in Quantum Biology \cite{ball2011physics,lloyd2011quantum, lambert2013quantum,buchleitner2014focus,mohseni2014quantum,salari2017quantum}, Vaziri \& Plenio proposed a quantum model for the KcsA selectivity filter \cite{vaziri2010quantum}. They compare ion-channel dimensions with the ion thermal wavelength to conjecture that the underlying dynamics for transport and selectivity might not be entirely classical. 
They propose a simplified model where  a single particle can undergo quantum tunneling on which the  interplay between quantum coherence and thermal noise would  be essential to the current efficiency. Besides that, they discuss a patch clamp experiment where their hypothesis could be tested. However, it is well known that Coulomb repulsion may deeply alter the tunneling rate \cite{imada1998metal}.
So here, we extend their quantum model to  explore one further aspect: Coulomb interaction in a two particle model with tunneling.

The paper is organized as follows. In Section 2, we estimate characteristic timescales for transport of repelling ions via perturbation theory and we build a two-particle model with Coulomb repulsion coupled to the environment via Lindblad operators. In Section 3 we show numerical results for ion transport, concluding that Coulomb repulsion dictates timescales and imposes bounds to  the free parameters of our model. Finally, we add  noise via dephasing and incoherent thermal hopping. We observe noise improves ion transport in a certain parameter range . This section is followed by the Conclusions.

\section{Model \& Methods}\label{section2}

The structure of the KcsA selectivity filter consists in four  binding sites for the ion. The axial separation between these potential minima is about $0.24$ nm and, based on detailed molecular dynamics simulation \cite{gwan2007cooperative}, the potential barrier height varies between $\sim 1.7-8.0 ~~ k_{B}T$ depending on both  protein  thermal vibrations and ion position in the channel. 
As proposed in \cite{vaziri2010quantum}, we assume a coherent process behind ion transmission, and we estimate the rates underlying the dynamics based on data from specialized literature   \cite{vaziri2010quantum, berneche2001energetics, kopfer2014ion, gwan2007cooperative}.

First, each ion crossing the channel finds a sequence of potential barriers formed by carbonyl groups. Approaching the potassium total energy by its thermal energy, $E=k_{B}T/2\approx 1.28\times 10^{-2}$ eV, and supposing a rectangular potential barrier, we can estimate a tunneling probability,
$p_{tun}\sim e^{-\Delta\sqrt{2m\Delta E/h^{2}}}$,
where $\Delta E$ is the difference between the barrier energy $1.7k_{B}T$ (the most favorable case) and the ion thermal energy. The barrier width, $\Delta$, is more difficult to estimate, but it is certainly  smaller than the $0.24$ nm distance between consecutive channel sites. Due to the exponential dependence of the transmission probability with barrier width, small $\Delta$ variations imply large probability variations. Also, depending on the energy  associated to the  minimum between two consecutive potential barriers, the ion kinetic energy, $K$, can vary from $\sim 0.5k_BT$ to $\sim 1.7k_BT$ \cite{kopfer2014ion}. We use this energy interval to estimate the ion frequency inside the well, $\nu=K/h$, and find the tunneling rate, 
$t_r\sim \nu~ p_{tun}$,  lying in the range $
t_r\sim 10^{11}-10^{13}\,\text{s}^{-1}$.

An experimental setup is suggested in Ref. \cite{vaziri2010quantum} where the ionic current of  $\sim\, 1\, \text{pA}$ could be possibly measured. But also, values of  the	order of $\sim 10\, \text{pA}$ are found in classical molecular simulations \cite{kopfer2014ion}. So we assume  ionic currents  varying from $1 \,\text{pA}$ to $10\,\text{pA}$, and estimate an ion rate  flowing through the channel of the order of $\sim 10^{7}-10^{8} \,\text{s}^{-1}$. 
 
In this work, we deal with a two-particle model where coherent tunneling and Coulomb repulsion play a central role in ion transmission. We start investigating  the relation between hopping rate in adjacent sites, $c$, and Coulomb repulsion $U$, using  second-order perturbation theory \cite{sakurai2014modern}. Assuming $ \hbar c\ll U$  yields an effective hopping 
$c_{eff} \approx  \hbar c^{2}/U$ as  the effective tunneling rate between adjacent sites. This effective hoping, $c_{eff}$, imposes different  timescales to the system as it will be shown later on. We associate the above measured ionic current to a maximum effective hopping rate, that is, $c_{eff} \sim 10^{7}-10^{8}\,s^{-1}$. Now, considering ions dehydrated of their water shielding we find,
\begin{equation}
U = \frac{ke^{2}}{a}\approx 5.14 \,\text{eV},
\label{uestima}
\end{equation}
where $a\sim 0.24$ nm is the distance between neighboring sites, $k$ is the Coulomb's constant and $e$ is the electron charge. We may then estimate the hopping from the electrostatic repulsion energy and the effective hopping,
\begin{equation}
 \frac{\hbar c^{2}}{5.14\,\,\text{eV}}\sim 10^{7}-10^{8}\,\text{s}^{-1},
\label{tax}
\end{equation}
what yields a value close to the estimated tunneling rate, 

\begin{equation}
 c \sim t_{r} \sim 10^{11}-10^{13}\,\text{s}^{-1}\;.
\label{cwrong}
\end{equation}
So, the relation between  Coulomb  and hopping terms is $U/ \hbar c \sim 10^3 - 10^5$.

\subsection{The quantum model}

Ion energy states inside the well may couple to vibrating energies of the carbonyl groups giving rise to dephasing and coherence loss \cite{vaziri2010quantum}.
Here we consider the simplest possible model for two particles with a simple site independent dephasing to account for these carbonyl oscillations through a specific non-unitary term in the dynamics. Another thermal effect is the possible  incoherent scattering caused by oscillations in the potential barrier height, which is  treated as another noise term as it is shown below.

Extending the model by Vaziri \& Plenio \cite{vaziri2010quantum} we consider a two particle system with Coulomb interaction in a one-dimensional lattice. We use a tight-binding  Hamiltonian for the linear chain of potential barriers, each ion in the SF may cross it through  coherent tunneling.  The Hamiltonian describing the unitary evolution  is given by 
\begin{equation}
  H= - \hbar
 \sum_{j=1}^{4} c \prt{{\sigma}_{j}^{\dag}{\sigma}_{j+1}+{\sigma}_{j+1}^{\dag}{\sigma}_{j}} +V,
\label{tight}
\end{equation}
where $\hbar$ is the Planck's constant, $j$ is the  site label, $c$ is the hopping rate between adjacent sites.  ${\sigma}_{j}^{\dag}$ (${\sigma}_{j}$) is the creation (annihilation) operator for fermions in $j$th-site.
The potential energy $V$ due to the Coulomb repulsion is given  by

\begin{equation}
  V=\frac{1}{2}\sum_{j=1}^{4}\sum_{j'\neq j=1}^{4}\frac{U}{|j-j'|}\sigma_{j}^{\dag}\sigma_{j}\sigma_{j'}^{\dag}\sigma_{j'}.
\label{binding}
\end{equation} 

To simulate the SF exchange of ions with the environment, extra sites  are linked on each side of the chain \cite{plenio2008dephasing,toymodel}. 
A site labeled  site-$0$ acting as an external {\it source} of particles,
is connected with site-1, the first internal SF site.  The second extra site, site-6, is connect with the SF site-5 and  acts as a {\it drain}. Its role is to  remove ions from the channel, so that the drain is populated via irreversible decay. It is worth emphasizing  that  source and drain here are poorly compared to thermal baths, since they only act supplying (source) and leaking (drain) particles to/from the channel. They do introduce decoherence into the system as we discuss later on,  but  terms ruling the interaction of a system  coupled to thermal baths are indeed not considered in our model.

This approach is formalized by using the following Lindblad operators \cite{breuer2002theory}:

\begin{equation}
\mathcal{L}_{s}\prt{\rho}=\Gamma_{s}\prt{-\cha{{\sigma}_{0}^{\dag}{\sigma}_{1}{\sigma}_{1}^{\dag}{\sigma}_{0},\rho}+2{\sigma}_{1}^{\dag}{\sigma}_{0}\rho{\sigma}_{0}^{\dag}{\sigma}_{1}},
\label{lsource}
 \end{equation}where $s$ stands for source,  $\Gamma_{s}$ denotes source supplying rate and 
$\rho=\rho(t)$ is the density matrix (DM). Similarly to Eq.(\ref{lsource}), the correspondent Lindblad operator for the drain is given by

\begin{equation}
\mathcal{L}_{d}\prt{\rho}=\Gamma_{d}\Biggl(-\cha{{\sigma}_{5}^{\dag}{\sigma}_{6}{\sigma}_{6}^{\dag}{\sigma}_{5},\rho}+2{{\sigma}_{6}^{\dag}{\sigma}_{5}  \rho{\sigma}_{5}^{\dag}{\sigma}_{6}}\Biggr),
\label{ldrain}
\end{equation}
where $d$ stands for drain and  $\Gamma_{d}$ denotes the leaking rate of the drain. 
 
The drain is populated as the ions are transferred along the channel. One can then investigate the rate at which the drain is populated in order to study the channel transport efficiency. In this work we concentrate in exploring the role of Coulomb interaction among  ions inside the channel and neglect external fluctuations. At this point  Lindblad formalism comes at hand since it  implicitly assumes Markovian processes for drain and source. 

In this context it is natural to use  a fermion number occupation basis for the ion-channel sites (1 to 5), avoiding two particles in the same site. On the other hand, source and drain are treated in a boson occupation  basis, since it is irrelevant to the dynamics inside the channel. This results in a density matrix composed of 23 squared terms.

Still following Ref. \cite{vaziri2010quantum},  we model carbonyl group thermal vibrations adding a noise term that randomizes the local phase excitations at a rate $\gamma_{deph}$. Thus, we will assume a model where  noise is local and use the following Lindblad superoperator, 

\begin{equation}
\mathcal{L}_{deph}\prt{\rho}=\sum_{j=1}^5 \gamma_{deph}\prt{-\cha{{\sigma}_{j}^{\dag}\sigma_{j},\rho}+2{\sigma_{j}^{\dag}\sigma_{j}\rho{\sigma}_{j}^{\dag}{\sigma}_{j}}}\;.
\label{ldep}
 \end{equation}
 This dephasing   produces an exponential decay of the density matrix terms with rate $\gamma_{deph}$ \cite{contreras2014dephasing}. Besides the random phase, we also consider another term taking into account particles  jumping incoherently due to thermal excitations. This incoherent scattering also produces an exponential decay of the DM elements  implying coherence loss.  It is given by

\begin{equation}
\begin{split}
\mathcal{L}_{th}\prt{\rho}&=\sum_{j=1}^{4}\Gamma_{th}\cch{2\sigma_{j+1}^{\dag}\sigma_{j}\rho \sigma_{j}^{\dag}\sigma_{j+1}-\cha{\sigma_{j}^{\dag}\sigma_{j+1}\sigma_{j+1}^{\dag}\sigma_{j},\rho}}\\&+ \sum_{j=1}^{4}\Gamma_{th}\cch{2\sigma_{j+1}\sigma_{j}^{\dag}\rho \sigma_{j}\sigma_{j+1}^{\dag}-\cha{\sigma_{j}\sigma_{j+1}^{\dag}\sigma_{j+1}\sigma_{j}^{\dag},\rho}},
\end{split}
\label{thermal}
\end{equation}where  $\Gamma_{th}$  is the thermal rate.

Thus, the evolution of the system is obtained by the integration of the following Lindblad master equation \cite{breuer2002theory,rivas2012open}

 \begin{equation}
 \frac{\mathrm{d}}{\mathrm{d} t}{\rho}\prt{t}= -\frac{i}{\hbar}\cch{H,\rho\prt{t}}+\mathcal{L}_{s}\prt{\rho}+\mathcal{L}_{d}\prt{\rho}+\mathcal{L}_{deph}\prt{\rho}+\mathcal{L}_{th}\prt{\rho}.
\label{lidbladdiago2}
\end{equation}
which defines a set of  coupled linear
ordinary differential equations.

\section{{Numerical  results}}

 The numerical results  presented here are conveniently   rescaled by the effective hopping,  $c_{eff}\approx \hbar c^2/U$. Adimensional time is given by $\tau=t\,c_{eff}$ and system rates $\Gamma_s$, $\Gamma_d$,  $\gamma_{deph}$ and $\Gamma_{th}$ are rescaled dividing by  $c_{eff}$, e.g, $\tilde{\Gamma}_s=\Gamma_s/c_{eff}$. 
From Equation (\ref{cwrong}) we assume $c$  in the range $[10^{11} -10^{13} ]s^{-1}$, which combined with Equation \ref{uestima} results in an adimensional Coulomb energy $~{\cal U} = U /\hbar c~$  in the range $[10^3 - 10^5]$.  We set $\hbar=c=1$, so ${\cal U} = U$. Also, the initial condition is two particles at site-0, the source.

\begin{figure}[ht]
\includegraphics[width=0.8\linewidth, clip=true]{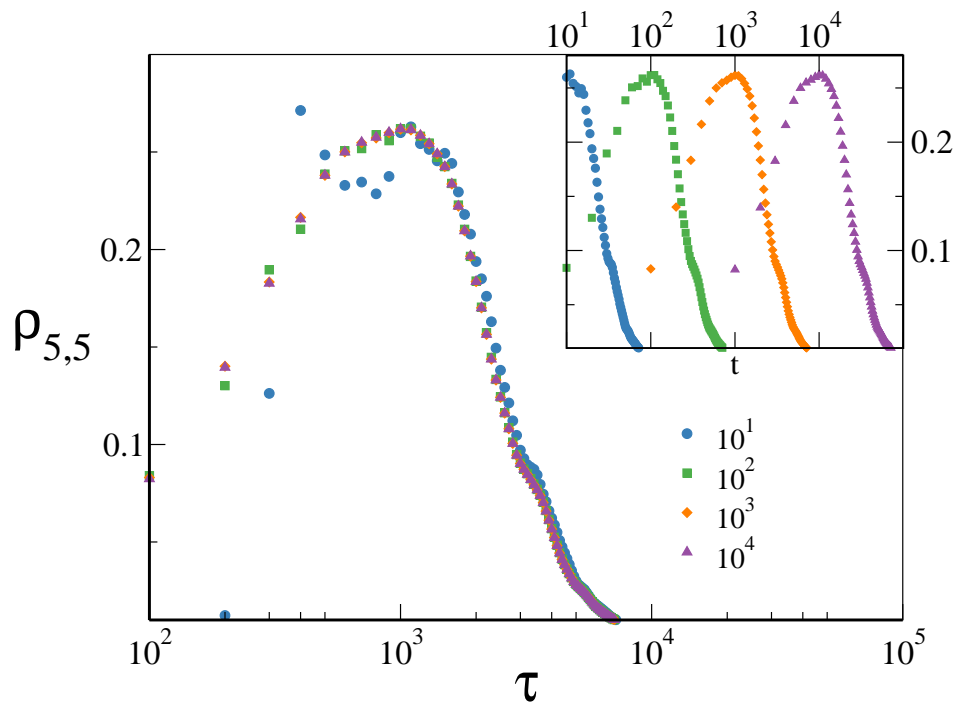}
\caption{(Color online) Site-5 population versus reduced time (logarithmic scale in $\tau$-axis) for different values of Coulomb strength ${U}$ for  $\tilde{\Gamma_{s}}=\tilde{\Gamma_{d}}=1$ and ($\tilde{\gamma}_{deph}=\tilde{\Gamma}_{th}=0$). Raw data are shown in the inset with the same convention. The curves matching shows the time rescaling with  Coulomb repulsion. }
\label{sitio5rescaledifU}
\end{figure}

As discussed above in section \ref{section2}, different $ U$ values imply different system timescales. Fig. (\ref{sitio5rescaledifU}) shows the  site-5 population for different ${U}$ values as function of the reduced time.
Recall that site-5 population determines the  drain occupation, since the Lindblad operator connects them trough Equation (\ref{ldrain}). The inset in Fig. (\ref{sitio5rescaledifU})  shows  these curves without time rescaling. The larger the  Coulomb repulsion, the slower the dynamics.  

Since the same dynamics is obtained if we correctly rescale time using Coulomb repulsion,  we may set a constant value for 
 ${U}$. So, for the remaining results in this paper we set ${U}=10^3$, which is the lower bound of the interval estimated above for $U$. We  investigate then different scenarios generated by constraints imposed by external rates. In particular, we search an optimal ion flow in the parameter space generated by these different rates.

\begin{figure}[h!]
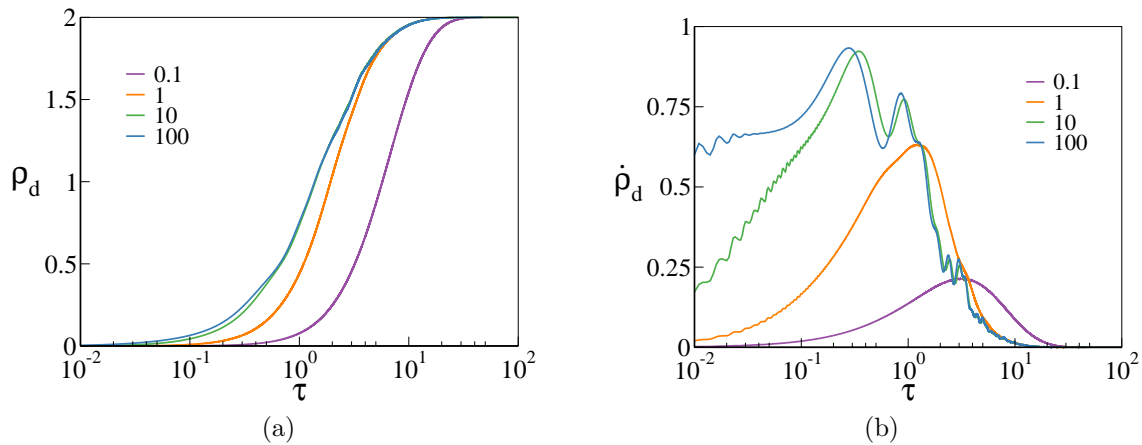

\centering
\subfloat[]{{\includegraphics[width=7cm,clip=true]{draindifgs.eps} }}%
\qquad
\subfloat[]{{\includegraphics[width=7cm,clip=true]{derivdraindifgs.eps} }}%
\caption{(Color online) (a) Drain population $\rho_d$ and  (b) its time derivative $\dot{\rho}_d$  for the source rate $\tilde{\Gamma}_{s}$ indicated. Drain rate $\tilde{\Gamma}_d=1$. }
\label{derivdifgs}
\end{figure}

Fig. \ref{derivdifgs}(a) shows  drain  population, $\rho_d\equiv\rho_{6,6}$, for different  $\tilde{\Gamma}_{s}$ values and $\tilde{\Gamma}_{d}=1$. At short times ($\tau \lesssim 1$) the drain is populated faster for $\tilde{\Gamma}_{s}>1$. If $\tilde{\Gamma}_{s}=1$ there is a slower dynamics at short times but at longer times it grows faster and the drain is fulfilled at the same time than for $\tilde{\Gamma}_{s}>1$.    Fig. \ref{derivdifgs}(b)  shows the time derivative of the drain population. The high frequency oscillations dominating at short times are related to the hopping $c$, since there is  a single particle in the channel. The subsequent low frequencies are in the typical time scale of the effective hopping, since now two particles partially coexist in the channel.  The later is more pronounced when particle injection is faster than $\tilde{\Gamma}_s>1$.  

\begin{figure}[h!]
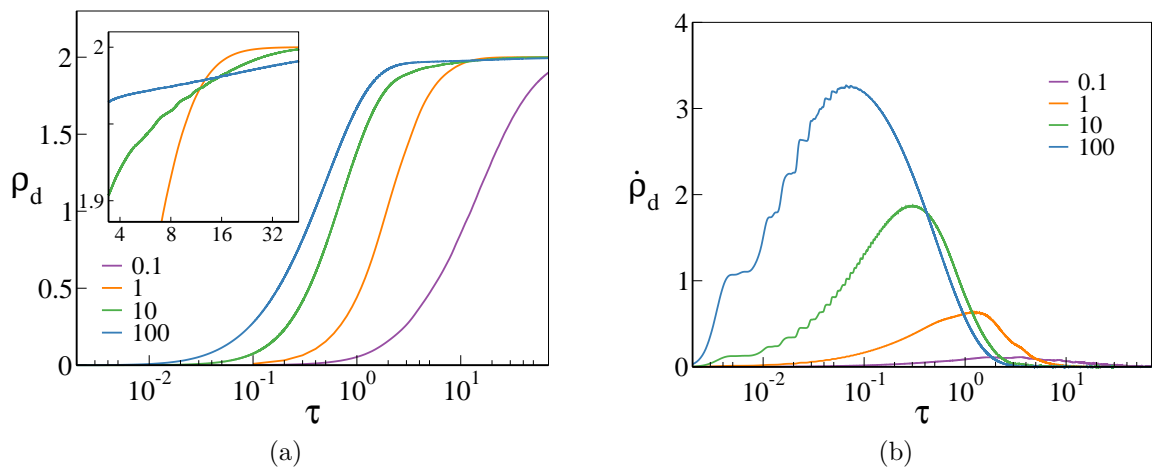

\centering
\subfloat[]{{\includegraphics[width=7.2cm,clip=true]{draindifgd.eps} }}%
\qquad
\subfloat[]{{\includegraphics[width=7cm,clip=true]{derivdraindifgd.eps} }}%
\caption{(Color online)
(a) Drain population $\rho_d$ and (b) its time derivative $\dot{\rho}_d$ for  the drain rate $\tilde{\Gamma}_{d}$ indicated. Source rate $\tilde{\Gamma}_s=1$.}
\label{draindifgd}
\end{figure}

\begin{figure}[h!]
\centering
\includegraphics[width=7cm, clip=true]{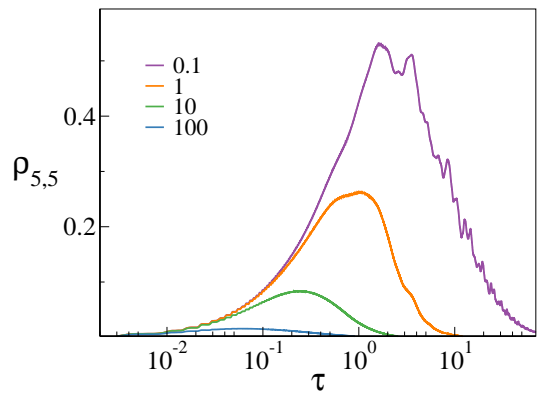}
\caption{(Color online) Site 5-occupation versus time for indicated drain rates $\tilde{\Gamma}_{d}$ and source rate $\tilde{\Gamma}_s=1$. }
\label{U1000sitio5dif}
\end{figure}

Now we explore cases varying drain rates at constant input. Fig. \ref{draindifgd}(a) illustrates the case. The drain is occupied faster when $\tilde{\Gamma}_{d} = 1$ (inset in Fig. \ref{draindifgd}(a)). When drain occupation is asymptotically close to two particles, dynamics slows down for  $\tilde{\Gamma}_{d}>1$.    Similarly to Fig. \ref{derivdifgs}(b) we observe in Fig. \ref{draindifgd}(b) high frequency oscillations associated to single particle hopping, $c$. In Fig. \ref{draindifgd}(b) the drain is  occupied fast for  larger values of $\tilde{\Gamma}_d$ up to the time when site-5 occupation is maximum. After that time, larger values of $\tilde{\Gamma}_{d}$ imply depletion of site-5 occupation (Fig. \ref{U1000sitio5dif}). 

\section{Noise-assisted Transport}

The results presented in the last section for ion transport properties referred to a noiseless system. Now, we expect that a certain amount of noise may wipe out quantum  coherence. So, in what follows we analyze the impact  of dephasing and thermal noises on ion transport.

\begin{figure}[h]
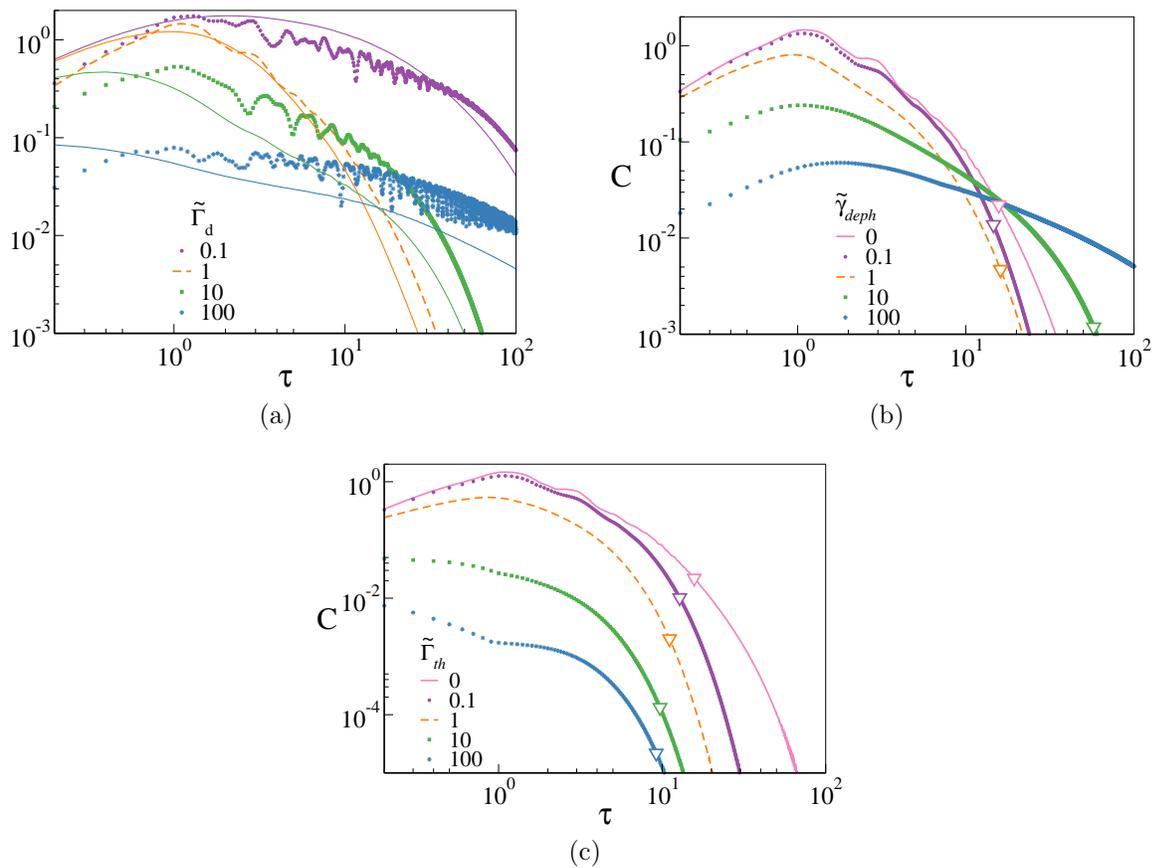

\centering
\subfloat[]{{\includegraphics[width=7cm,clip=true]{deco.eps} }}%
\qquad
\subfloat[]{{\includegraphics[width=7.2cm,clip=true]{coerenciadephasing.eps} }}%
\\
\subfloat[]
{{\includegraphics[width=7cm,clip=true]{coerenciathermal.eps} }}%

\caption{(Color online)(a) Coherence evolution (non-continuous lines) and occupation inside the channel (continuous lines) for different $\tilde{\Gamma}_{d}$ values at constant $\tilde{\Gamma}_{s}=1$. (b) Coherence evolution for different $\tilde{\gamma}_{deph}$ values, drain occupation limit of 1.99 for $\tilde{\gamma}_{deph}=10$ is reached for $\tau>300$ (not shown).  (c) different $\tilde{\Gamma}_{th}$  at constant $\tilde{\Gamma}_{s}=\tilde{\Gamma}_d=1$, down triangles indicate when  $\rho_{d}=1.99$.}
\label{coerenciathermal}
\end{figure}

Coherence is investigated studying the off-diagonal terms of the  density matrix reduced to the channel sites. For that we use  the l$_{1}$-norm defined as \cite{baumgratz2014quantifying}  

\begin{equation}
C(\tau)= \sum_{i\neq j} \mid \rho_{ij}(\tau) \mid.
\end{equation}
In our simulations the system is considered coherent for times up to the order of $\tau$, corresponding to real times of the order of  $10^{-8}$s \cite{vaziri2010quantum}.  
Fig. \ref{coerenciathermal}(a) shows $C(\tau)$ at constant $\tilde{\Gamma}_s=1$ for different $\tilde{\Gamma}_d$ values (non-continuous lines, $\tilde{\Gamma}_d$ values indicated in the figure).  The channel population respective to each $\tilde{\Gamma}_d$ case is shown by the solid lines. 
Clearly $C(\tau)$ follows the same behavior of the of  remaining occupation inside the channel. Also, as above in Fig. \ref{draindifgd}(a), drain is more quickly filled up when $\tilde{\Gamma}_d=1$. In Figs. \ref{coerenciathermal}(b) and \ref{coerenciathermal}(c) triangles indicate 1.99 as particle occupation in the drain, or $0.01$ still in the channel. We have arbitrarily chosen also this as the minimum value to consider a system coherent, that is,
 there is coherence if $C(\tau)\gtrsim 10^{-2}$.

Fig. \ref{coerenciathermal}(b) shows the coherence $C(\tau)$   when $\tilde{\Gamma}_s=\tilde{\Gamma_d}=1$ for different values of $\tilde{\gamma}_{deph}$. We observe that a dephasing noise in the order of the system characteristic timescale ($\tilde{\gamma}_{deph}\lesssim 1$) favors drain-filling. Note the occupation triangles indicating a faster drain filling for these cases. 
 
Fig. \ref{coerenciathermal}(c) shows that thermal  noise always reduces drain-filling time. However  coherence loss is more pronounced than with noise dephasing, at least for $\tilde{\Gamma}_{th} > 1$. The occupation value $10^{-2}$ (triangle in the figure) for $\tilde{\Gamma}_{th}=100$, for example, happens for $C(\tau)\sim 10^{-5}$. So the system is not considered coherent any longer.
On the other hand, for $\tilde{\Gamma}_{th}\leq 1$ coherence is preserved and filling drain times are smaller than in the noiseless system. 

\section{Conclusions}

Previous single particle quantum models \cite{vaziri2010quantum} suggest that tunneling rates in the order of $10^8 s^{-1}$ may explain ion channel high efficiency transport. However such models ignore the  presence of a large Coulomb interaction, which has been shown to display a relevant role in classical approaches \cite{bordin2012ion,kopfer2014ion}. So, here we improve the model considering a two-particle system with Coulomb repulsion.
This interaction is known to  slow down the dynamics, particularly when the interaction is strong, which is the case here. Small changes in the estimated values for the  potential barrier may alter hopping characteristic scales to ranges $[10^{11}-10^{13}]s^{-1}$, around $10^3$ times faster than previously estimated \cite{vaziri2010quantum}. This results in an effective hopping term still compatible with values expected for the channel current, while keeping a coherent quantum system. 

Our simulations show that ion transport is optimized if source input  and drain output rates are compatible with the system characteristic time, as defined by the relation $c_{eff}=\hbar c^2/U$.  In fact, when input/output rates are correctly adjusted,  Coulomb repulsion is minimized. Considering  larger rates, e.g., $\tilde{\Gamma}_s>1$ or $\tilde{\Gamma}_d>1$, results in low ion currents and a large dwell time for ions inside the channel. This remaining charge oscillates with the system characteristic time, $\tau$, as indicates the current derivative.  

In order to check the possibility of improving charge transport with noise, as in Ref. \cite{vaziri2010quantum}, we varied dephasing rate and thermal noise over  four orders of magnitude. Simulations show that dephasing noise levels in the order of one (that is, in the order of the effective hoping) or smaller enhance charge transport. Thermal noise increases ion transport independently of its intensity, however, coherence is lost for values above one.

Lindblad  operators artificially impose input/output rates to emulate ion concentrations outside the channel. Our results show that  Coulomb repulsion dictates a characteristic timescale for the ion transport inside the channel. If ion concentration is below some optimal value, we expect a high speed transport since the problem is reduced to a single particle one with a hopping one thousand times faster. So, at these low densities we expect a linear relation between current and concentration with a high slope (e.g. concentrations below 10mM in Ref. \cite{morais2001energetic}).  On the other side, if ion concentration is above a single particle regime, dynamics slows down and we also expect  a linear behavior but with a smaller slope. However,  below 10mM many potassium channels may undergo inactivation what makes this concentration interval delicate for experiments \cite{zhou2001chemistry}.

\ack 
We thank the Brazilian agencies CAPES, CNPq, and
FAPERGS for financial support. We also acknowledge the Computational Center of the Physics Institute of UFRGS for cluster computing hours.

\bibliography{bib}                 
\end{document}